\documentstyle[prd,preprint,aps,epsfig]{revtex}

\def\bea {\begin{eqnarray}}
\def\eea {\end{eqnarray}}
\def\be {\begin{equation}}
\def\ee {\end{equation}}
\def\ben{\begin{enumerate}}
\def\een{\end{enumerate}}
\def\bi{\begin{itemize}}
\def\ei{\end{itemize}}
\def\ie{{\it i.e.}}

\def\etal{{\it et al.}}
\def\prl {Phys. Rev. Lett.\ }

\def\pr {Phys. Rev.\ }
\def\np {Nucl. Phys.\ }

\begin{document} 
\tighten
\draft 
\preprint{ } 
\title{$^{8}$Li electron spectrum versus $^{8}$B neutrino spectrum:
implications for the Sudbury Neutrino Observatory}
\author{ G. Jonkmans$^{1,2}$\footnote{Present address: 
Institut de Physique, Universit\'e de Neuch\^atel, CH-2000 Neuch\^atel,
Switzerland},
I.S. Towner$^{1}$ and B. Sur$^{2}$} 
\address{$^{1}$ Department of Physics, Queen's University, Kingston,
Ontario, Canada  K7L 3N6}
\address{$^{2}$ AECL, Chalk River Laboratories, Chalk River,
Ontario, Canada  K0J 1J0}
\date{\today} 
\maketitle
\begin{abstract} 

The sensitivity of the Sudbury Neutrino Observatory (SNO) to measure the
shape of the recoil electron spectrum in the charged-current reaction
of $^{8}$B solar neutrinos interacting with deuterium can be improved
if the results of a $^{8}$Li beta-decay calibration 
experiment are included in the test.  We calculate an improvement
in sensitivity, under certain idealistic assumptions, of about 
a factor of 2, sufficient to resolve different
neutrino-oscillation solutions to the solar-neutrino problem.
We further examine the role of recoil and radiative corrections on
both the $^{8}$B neutrino spectrum and the $^{8}$Li electron spectrum
and conclude that the influence of these effects on the 
ratio of the two spectra as measured by SNO is
very small.

\end{abstract} 

\pacs{13.15.-f, 23.40.Bw, 25.30.Pt, 96.60.Kx}

\narrowtext 

\section{Introduction}
\label{s:intro}

The Sudbury Neutrino Observatory (SNO) \cite{SNO} will utilize the
interaction of $^{8}$B solar neutrinos with deuterium in heavy water to
measure the shape of the recoil electron spectrum in charged-current (CC)
interactions and the ratio of the number of charged-current to
neutral-current (NC) events. 
In what follows, we 
will concentrate only on the test of the CC spectral 
shape, where one of the measurable quantities
is the average kinetic energy of the detected recoil 
electrons, $ \langle T_e
\rangle_{\nu}$, to be defined more precisely below.  
The main uncertainties,
other than counting statistics, arise from uncertainties in (a) the 
theoretical standard $^{8}$B neutrino spectrum, (b) the 
detector energy resolution and (c) the detector absolute energy scale.  
If the uncertainties from these sources can be reduced sufficiently
SNO should be able to distinguish between a Standard Solar Model with
no neutrino oscillations and one with various choices of
neutrino oscillation scenarios \cite{BA89}: a Small Mixing Angle (SMA),
a Large Mixing Angle (LMA) or Vacuum Oscillations (VAC) solutions.
One way to reduce these uncertainties is to introduce a $^{8}$Li
calibration source. 
As part of its overall calibration strategy, the SNO collaboration
will install a system that is capable of producing a $^{8}$Li source
placed at several different locations inside the detector \cite{Su94}.
Detection of electrons from this source would demonstrate
that the results obtained by SNO for a known beta-decay spectrum are
consistent with those measured in the laboratory.  

In this paper we point out that uncertainties in the measured CC
spectral shape can be reduced considerably by a direct comparison
with the measured beta-decay spectrum of $^{8}$Li.  For instance,
rather than considering
$\langle T_e \rangle_{\nu}$ to be the benchmark for comparing experiment
with different theoretical expectations, we consider the ratio
$\langle T_e \rangle_{\nu} / \langle T_e \rangle_{e}$, where
$\langle T_e \rangle_{e}$ is the average kinetic energy of detected 
electrons from $^{8}$Li decay.  In forming ratios such as this,
many uncertainties cancel producing a much improved error budget.
This suggested analysis strategy is based upon the fact that the electron 
spectra from $^{8}$Li
beta decay and from $^{8}$B solar-neutrino absorption on deuterium 
originate from mirror weak decays to common final 
states in $^{8}$Be$^*$, hence are highly correlated,
have somewhat similar shapes and cover essentially the same 
energy range.  In formulating this strategy, we have calculated 
the forbidden and radiative corrections to the $^{8}$Li beta 
and $^{8}$B neutrino spectra and have considered the effects of 
theoretical and experimental uncertainties on 
the above-mentioned ratio.

In Sec.\ \ref{s:s2} we define more carefully the quantities involved
and outline the suggested strategy.  
In Sec.\ \ref{s:s3} we give some calculated
results, and in Sec.\ \ref{s:s4} we summarize our findings.  Most of the
formulae have been relegated to appendices. 

\section{The Measured Quantities}
\label{s:s2}

The recoil electrons produced by $^{8}$B solar neutrinos being absorbed
on deuterium have an average kinetic energy (for an ideal detector)
given by

\be
\langle T_e \rangle_{\nu}^{{\rm ideal}} = \frac
{ \int_{T_{\rm min}} dT_e T_e \int dE_{\nu} \lambda_{\nu}(E_{\nu})
P_{ee}(E_{\nu}) \frac{d \sigma_{CC}}{dT_e} (E_{\nu}) } 
{ \int_{T_{\rm min}} dT_e  \int dE_{\nu} \lambda_{\nu}(E_{\nu})
P_{ee}(E_{\nu}) \frac{d \sigma_{CC}}{dT_e} (E_{\nu}) } , 
\label{Tenuideal}
\ee

\noindent where $T_{\rm min}$ is the threshold kinetic energy below which
events in SNO will be discarded.  For the discussions that follow,
we adopt $T_{\rm min} = 5$ MeV.  The charge-current differential
cross-section for neutrino absorption on deuterium $d \sigma_{CC} / dT_e$,
is a function of neutrino energy.  We will use the calculated
cross-section of Ellis and Bahcall \cite{EB68} for our computations,
and adopt from Bahcall and Lisi \cite{BL96} a 0.14 \% $1 \sigma$
error in $\langle T_e \rangle_{\nu}$, this being the difference when
Kubodera-Nozawa cross-sections \cite{KN94} are used instead.  The $^{8}$B
neutrino spectrum is denoted $\lambda_{\nu}(E_{\nu}) \equiv d\Gamma /
dE_{\nu}$, 
and $P_{ee}(E_{\nu})$
is the probability that an electron-neutrino produced at the core of the
sun remains an electron-neutrino by the time it is detected on earth.
The different neutrino-oscillation scenarios yield different survival
functions, $P_{ee}(E_{\nu})$.  In the Standard Model with no neutrino
oscillations, $P_{ee}(E_{\nu}) = 1$.

The decay electrons produced in $^{8}$Li beta decay have an average kinetic
energy (for an ideal detector) given by

\be
\langle T_e \rangle_{e}^{{\rm ideal}} = \frac
{\int_{T_{\rm min}} dT_e T_e \lambda_{e}(T_e)}
{\int_{T_{\rm min}} dT_e  \lambda_{e}(T_e)} ,
\label{Teeideal}
\ee

\noindent where $\lambda_{e}(T_e) \equiv d\Gamma /dE_e$ 
is the electron spectrum.  The beta decay of
$^{8}$Li is the isospin analogue of the beta decay of $^{8}$B; both
populate the same daughter $^{8}$Be states.  The $^{8}$Be excited states
are unstable and break up into two alpha particles.  As a consequence,
the shape of the $^{8}$Li $\beta^{-}$ spectrum and of the $^{8}$B
neutrino spectrum deviate significantly from the standard allowed 
shape.  Measurements \cite{Expts} of the 
delayed $\alpha$ spectra allow one to
determine the profile of the intermediate $^{8}$Be state and thus to
calculate the deviations in the electron and neutrino spectra.  This is
discussed at length in Bahcall \etal \cite{Ba96}, where it is 
shown there is
considerable discrepancy among different experiments related to the
absolute energy calibration of the alpha particles.  
Bahcall \etal \cite{Ba96}
choose to display this uncertainty as a possible offset $b$ in the energy 
of the alpha particles from $^{8}$Be break up: $E_{\alpha} \rightarrow
E_{\alpha} + b$.  The value of the offset affects the calculated shape
of {\it both} the beta and neutrino spectra.  
The effective $3 \sigma$ uncertainty
of the offset is estimated to 
be $b = \pm 0.104$ MeV \cite{Ba96}; this estimate
includes theoretical errors.  Bahcall \etal \cite{Ba96} have tabulated
their recommended $^{8}$B neutrino spectrum together with their $3 \sigma$
errors.

In this work we require both a $^{8}$B neutrino spectrum and a $^{8}$Li
electron spectrum from decay to the same final states 
in $^{8}$Be$^*$, in order to compare the effects of 
common-mode uncertainties. Therefore we use R-matrix
theory to model the profile of the intermediate $^{8}$Be state, and
in particular, the R-matrix fits of Barker \cite{Ba89}.  To reproduce
the possible offset in the energy of alpha particles, we shift the
energy of the dominant resonance, $E_1 \rightarrow
E_1 + b^{\prime}$, where the offset $b^{\prime}$ is 
obtained from the requirement
that this method gives the same error as the Bahcall \etal \cite{Ba96}
treatment.  The same offset is used consistently for both decay spectra.
Details of the R-matrix analysis are given in Appendix \ref{s:Rfit}.

There are two further differences between the $^{8}$B neutrino spectrum
and the $^{8}$Li electron spectrum originating in (a) recoil (or forbidden)
corrections and (b) radiative corrections.  Details of these have been
relegated to Appendices \ref{s:ARc} and \ref{s:ARdc} respectively.  
However, in Figs.\ \ref{f:1} and \ref{f:2} we show plots
of these corrections for the individual spectra and their ratio.  Although
both corrections are energy-dependent and hence a factor in determining
$\langle T_e \rangle$ their ratio is very much less energy dependent.
Thus any error in $\langle T_e \rangle$ associated with determining these
corrections is much reduced in considering the ratio
$\langle T_e \rangle_{\nu} / \langle T_e \rangle_{e}$.

In Fig.\ \ref{f:1} we show the recoil correction 
for the $^{8}$B neutrino spectrum
and $^{8}$Li electron spectrum for two cases.  The calculation is based
on the elementary-particle treatment of beta decay of Holstein \cite{Ho74}
and the correction depends principally on two parameters: $b/c$ and $d/c$.
Here $b$, $c$ and $d$ are the weak-magnetism, Gamow-Teller and
induced pseudotensor current form factors respectively.  
The two cases correspond to the
induced pseudotensor current being retained in or 
removed from the calculation.
In both cases the ratio is almost independent of energy, 
indicating little sensitivity 
to the presence of induced pseudotensor currents.

In Fig.\ \ref{f:2} we show the radiative correction for the $^{8}$B neutrino
spectrum and $^{8}$Li electron spectrum again for two cases.  In the 
first case, we assume the internal bremsstrahlung 
photons are not detected and so
the energy of the photon is integrated over in obtaining the
radiative correction. In this case there is a large difference 
in the energy dependence between the 
radiative corrections for the electron and neutrino spectra.  In the
second case, we account for the fact that in a calorimetric 
detector such as SNO, the internal bremsstrahlung photons from a
$^{8}$Li source placed in the detector would be recorded and 
summed with the beta energy.
We assume, for simplicity, that the efficiency and detector 
response for electrons
and photons is the same, and compute the radiative correction for  
the summed energy (photons plus electrons) deposited.  Now one sees
there is a much reduced difference between the energy 
dependence of the radiative corrections
for the electron and neutrino spectra.

From Figs.\ \ref{f:1} and \ref{f:2}, 
it is fairly clear there will be little uncertainty in
the ratio, 
$\langle T_e \rangle_{\nu} / \langle T_e \rangle_{e}$,
arising from any uncertainty in recoil and radiative corrections.
Therefore, we will not consider these corrections explicitly any
further; but rather we will assume they are effectively absorbed in the 
energy offset parameter, $b$, which as has been mentioned above was
estimated by Bahcall \etal \cite{Ba96} to be $\pm 104$ keV, and
contained theoretical uncertainties associated with recoil and
radiative corrections.

Finally, we turn to detector-related uncertainties.  
The measured electron 
kinetic energy, $T_e$, determined by SNO will
be distributed around the true energy, $T_e^{\prime}$, with a width
established by the detected Cerenkov photon statistics.  
The resolution function
$R(T_e^{\prime},T_e)$
is approximated by a normalized Gaussian:

\be
R(T_e^{\prime},T_e) = \frac{1}{\sigma ( 2 \pi )^{1/2}}
\exp \left \{ - \frac{(T_e^{\prime} - T_e )^2}{2 \sigma^{2}} \right \} ,
\label{resfn}
\ee

\noindent where $\sigma \equiv \sigma (T_e^{\prime})$ is an
energy-dependent $1 \sigma$ width given by \cite{BL96}

\be
\sigma (T_e^{\prime}) = \sigma_{10} \sqrt{ \frac{T_e^{\prime}}{10~{\rm MeV}}},
\label{sig10}
\ee

\noindent with $\sigma_{10}$ the resolution width at $T_e^{\prime} = 10$ MeV.
We follow Bahcall and Lisi \cite{BL96} and use
$\sigma_{10} = 1.1 \pm 0.11$ MeV ($1 \sigma$ errors) as an illustrative
example.

The expressions, Eqs.\ (\ref{Tenuideal}) and (\ref{Teeideal}), are now
modified to include a response function

\be
\langle T_e \rangle_{\nu} = \frac
{ \int_{T_{\rm min}} dT_e T_e \int dE_{\nu} \lambda_{\nu}(E_{\nu})
P_{ee}(E_{\nu}) \int dT_e^{\prime} R(T_e^{\prime},T_e) 
\frac{d \sigma_{CC}}{dT_e^{\prime}} (E_{\nu}) } 
{ \int_{T_{\rm min}} dT_e  \int dE_{\nu} \lambda_{\nu}(E_{\nu})
P_{ee}(E_{\nu}) \int dT_e^{\prime} R(T_e^{\prime},T_e) 
\frac{d \sigma_{CC}}{dT_e^{\prime}} (E_{\nu}) } , 
\label{Tenu}
\ee

\be
\langle T_e \rangle_{e} = \frac
{\int_{T_{\rm min}} dT_e T_e \int dT_e^{\prime} R(T_e^{\prime},T_e) 
\lambda_{e}(T_e^{\prime})}
{\int_{T_{\rm min}} dT_e  \int dT_e^{\prime} R(T_e^{\prime},T_e) 
\lambda_{e}(T_e^{\prime})} .
\label{Tee}
\ee

\noindent The other major uncertainty concerns the absolute energy 
calibration for the SNO detector.  For this, we assume a 1\% error at 10 MeV.
This can be easily implemented in the calculations by modifying the
energy-resolution function, Eq.\ (\ref{resfn}), with the replacement

\be
R(T_e^{\prime},T_e) \rightarrow R(T_e^{\prime} + \delta , T_e) ,
\label{resfnrepl}
\ee

\noindent where $\delta = \pm 100$ keV ($1 \sigma $ error).

The above methods for treating detector-related uncertainties 
are a rough parametrization of detailed considerations regarding 
the spatial and directional response to energy deposition 
in the SNO detector. Other sources of 
uncertainties which have not been taken into account in this 
work are (a) the energy dependence of the detector efficiency, 
(b) the statistical and systematic uncertainties associated 
with background subtraction and neutral-current event separation, 
(c) the uncertainties in the $^8$Li beta-spectrum (as 
measured in SNO) due to the source container, and (d) the position
dependence of the measured energy.  
Evaluation of the 
effects of these uncertainties will have to await detailed 
Monte-Carlo simulations and actual measurements of detector 
performance when SNO is operational.

\section{The Error Budget}
\label{s:s3}

Having listed the source of uncertainties, it remains to quantify
their contribution to the error budget for the detection of recoil
electrons following neutrino absorption on deuterium alone,
$\langle T_e \rangle_{\nu}$,
or in concert with a calibration experiment with a $^{8}$Li source 
in the detector,
$\langle T_e \rangle_{\nu} / \langle T_e \rangle_{e}$.
The results are given in Table \ref{t:1}.  The error budget
for the former case has been given by Bahcall and Lisi \cite{BL96}
and we follow their example.  The statistical error ($3 \sigma $) on
5000 CC events is estimated at 0.98\% for
$\langle T_e \rangle_{\nu}$.
In the calibration experiment there will be considerably more events
counted, so the statistical uncertainty in the ratio
$\langle T_e \rangle_{\nu} / \langle T_e \rangle_{e}$
will be dominated by the 5000 CC neutrino events:  we use the same
statistical error.  For the neutrino-absorption on deuterium,
the uncertainty in the absorption cross-section is common to both
$\langle T_e \rangle_{\nu}$ and
$\langle T_e \rangle_{\nu} / \langle T_e \rangle_{e}$,
so again we use the same error.  The uncertainty in the neutrino spectrum
from the beta decay of $^{8}$B is characterised by the offset
parameter, $b$, associated with the absolute energy calibration of
the $\alpha$-particles in the measured delayed $\alpha$-spectrum.
Upto differences associated with isospin-symmetry breaking, recoil
and radiative corrections, the same uncertainty 
occurs in the calibration experiment, so the error in the ratio
$\langle T_e \rangle_{\nu} / \langle T_e \rangle_{e}$
is much reduced.  Table \ref{t:1} shows a 
reduction of uncertainty by a factor of 7.6.
The same reasoning follows for the detector-related uncertainties,
since in both experiments, energetic electrons are being counted in the
same detector.  We find a reduction in uncertainty by a factor of 8.5
due to the detector's energy resolution, and a factor of 5.0 due to its
absolute energy calibration.
The energy calibration of the detector here has been
oversimplified.  The analysis above assumes that the electron
sources from $^{8}$Li beta decay and from neutrino absorption
are identical, but of course they are not.  The $^{8}$Li
calibration sources are highly localized in space and time,
while the recoil electrons produced by absorption of solar
neutrinos are distributed throughout the detector and
distributed in time.  So our analysis here represents an optimal
scenario, but it is sufficient to make the point that an
improvement in the error budget will be achieved by performing
an in situ calibration experiment.

If all errors are added together in quadrature, then the $3 \sigma $
uncertainty in the CC test at SNO using
$\langle T_e \rangle_{\nu}$
alone is 2.8\% and is dominated by the uncertainty in the absolute
energy calibration.  However, in combination with a $^{8}$Li calibration
experiment, the $3 \sigma$ uncertainty in the ratio
$\langle T_e \rangle_{\nu} / \langle T_e \rangle_{e}$
is reduced to 1.2\%, and more importantly, is dominated by counting
statistics.  This analysis strategy will significantly improve
SNO's ability to discriminate among the different neutrino-oscillation
scenarios.  In Fig.\ \ref{f:3} we display the error budget together
with the theoretical value of
$\langle T_e \rangle_{\nu} / \langle T_e \rangle_{e}$
for various neutrino-oscillation solutions \cite{BK96} to the solar-neutrino
problem: the two (best-fit) Mikheyev-Smirnov-Wolfenstein (MSW)
solutions at small and large mixing angle (SMA and LMA) and the
purely vacuum (VAC) oscillation solution.  In an analagous plot
by Bahcall and Lisi \cite{BL96}, it is shown that a measurement of
$\langle T_e \rangle_{\nu}$
alone is unlikely to resolve the two MSW solutions.  On the other hand,
the ratio measurement, Fig.\ \ref{f:3}, clearly shows the LMA and SMA
solutions as being experimentally distinguishable, if the other 
uncertainties in the detector can be minimized.

Although we have considered the ratio of
$\langle T_e \rangle_{\nu} / \langle T_e \rangle_{e}$, it is possible
to construct other experimental quantities from the measured CC
and $^{8}$Li spectra. One of the simplest observables to consider is
the actual ratio of the normalized 
measured CC spectrum, $\lambda^{CC}_{exp}$,
to the normalized measured $^{8}$Li electron 
spectrum, $\lambda^{Li}_{exp}$.
Because of the high degree of similarity between those two spectra,
the effect of uncertainties on the detector response function are
expected to largely cancel when forming this ratio. 
In Fig.\ \ref{f:4} we display the effect of the uncertainty 
in the detector energy scale,
$\delta$ on the the double ratio

\be
R(CC/^{8}Li) = \frac{\lambda^{CC}_{exp}/\lambda^{CC}_{theor}}
{\lambda^{Li}_{exp}/\lambda^{Li}_{theor}}
\label{dblratio}
\ee

\noindent where $\lambda_{theor}$ refers to a spectrum 
where all parameters including the absolute
energy calibration are assumed to be known.  
If $\lambda^{Li}_{exp}$ is not measured (i.e. the denominator 
is assumed to be unity), then the dashed lines enclose the 
error band with $\pm$3$\sigma$ errors on $\delta$.  A 
measurement of $\lambda^{Li}_{exp}$ allows the cancellation of 
correlated errors and produces the much 
reduced ``$\pm$3$\sigma$'' error band enclosed by the solid 
lines.  The effect due to the statistics of 5000 CC 
events above 5 MeV is shown by 
overlaying on Fig.\ \ref{f:4} the shape distortion expected
for the best-fit mass and mixing values for the SMA solution \cite{BK96}
($\Delta m^{2} = 5.4 \times 10^{-6}$ eV$^{2}$ and sin$^{2} 2\theta =
7.9 \times 10^{-3}$).  We can clearly see that making use of the
$^{8}$Li spectrum as measured in SNO reduces the effects of
systematic uncertainties and greatly increases the discriminating power
of the CC shape measurement.

\section{Summary}
\label{s:s4}

The SNO detector is being constructed primarily to measure the
neutral-current (NC) to charged-current (CC) cross-section
ratio for solar neutrino absorption in deuterium.  It is
anticipated this measurement will give a strong and unambiguous
signal of neutrino oscillations (or the lack thereof).
Another test, which does not rely on the
knowledge of the absolute reaction cross-sections, is a
measurement of the spectral shape of the CC reaction alone.
To improve the sensitivity for this measurement, a
calibration experiment is planned at SNO with a $^{8}$Li source
placed in the detector.  If instead of considering only the first 
moment of the recoil electron spectrum in the CC reaction,
$\langle T_e \rangle_{\nu}$, the ratio of this moment
to the equivalent moment in the $^{8}$Li beta decay
is used as a benchmark, then an increase in the sensitivity of a factor
of 2.4 could be achieved in the CC-shape test. This improvement
would be sufficient to resolve the small-angle and large-angle MSW
solutions to the solar-neutrino problem.
We stress again, however, that not all uncertainties are
known at the present time and some will require actual
measurements when SNO is operational, so our analysis here
represents an optimistic best-case scenario.  To us, a
calibration experiment with a $^{8}$Li source certainly seems
beneficial.

\section{Acknowledgement}
\label{s:Ack}

We thank Emanuel Bonvin for initially suggesting the use 
of $^{8}$Li to calibrate the SNO detector, and Graham Lee-Whiting for
fruitful discussions on radiative corrections.
We also thank J.N. Bahcall and E. Lisi for permission to use their computer
code for the calculation of the neutrino-deuterium cross sections.

\appendix
\section{R-matrix fit}
\label{s:Rfit}

The allowed $\beta$-transitions from the decay of $^{8}$Li or $^{8}$B
populate $2^{+}$ states of $^{8}$Be, which then decay into two
$\alpha$-particles.  The $\alpha$-spectra
show a pronounced peak, corresponding to the $2^{+}$ first excited state
of $^{8}$Be at $E_x \simeq
3.0$ MeV.  Attempts to fit the $\alpha$-spectra
assuming only one state in $^{8}$Be is operative fail to give enough
yield at high energies.  Barker \cite{Ba69} proposed R-matrix formulae
in the many-level, one-channel approximation and used them to fit
the $\beta$-delayed $\alpha$-spectra.  More recent fits have been
given by Warburton \cite{Wa86} and Barker \cite{Ba89}.

In the many-level, one-channel approximation the beta-decay differential
cross-section is written

\be
d \Gamma \propto \frac{1}{\pi} g(E_x) p_e E_e E_{\nu}^2 F( \pm Z,E_e)
\delta (Q_{ec} + m - E_x - E_e -E_{\nu}) dE_x dE_e dE_{\nu} ,
\label{rate}
\ee

\noindent where $Q_{ec} + m = M - M^{\prime}$, with $M$ the 
mass of the parent
nucleus, $M^{\prime}$ the ground-state mass of the daughter nucleus
and $m$ the electron mass.  Here $E_x$ is the excitation energy in the
daughter nucleus, $E_e$ the electron energy, $p_e$ the electron momentum
and $F(Z,E_e)$ the usual Fermi function.  This expression differs from the
standard one by the presence of the function $g(E_x)/\pi$ and an
additional integration, $dE_x$, over the excitation energy of the
daughter nucleus.

The function $g(E_x)$ in R-matrix theory is

\bea
g(E_x) & = & \frac {\mid R_2(E_x) \mid^2 P_2(E_x) }
{ \mid 1 - \left ( S_2(E_x) - B_2 + {\rm i} P_2(E_x) \right ) 
R(E_x) \mid^2 } ,
\nonumber \\
R_2(E_x) & = & \sum_{\lambda} \frac {\gamma_{\lambda} (M_{GT})_{\lambda}}
{E_{\lambda} - E } ,
\nonumber \\
R(E_x) & = & \sum_{\lambda} \frac {\gamma_{\lambda}^2 }
{E_{\lambda} - E } ,
\label{Rfunctions}
\eea

\noindent where $E$ is the channel energy, \ie \ the energy above the
$\alpha + \alpha$ threshold: $ E = E_x + 0.092$ MeV.  Here $P_2(E)$
and $S_2(E)$ are the penetrability and shift factors for $L=2$ partial
waves and are expressed in terms of Coulomb functions for 
$\alpha + \alpha$ scattering evaluated at a chosen channel radius,
$a_2$.  Finally, $B_2$ is the boundary condition parameter.
The sum over $\lambda$ is a sum over all resonances retained in the
calculation.  For each resonance there are three parameters: $E_{\lambda}$
the centroid energy of the resonance, $\gamma_{\lambda}$ the reduced width 
amplitude for $\alpha + \alpha$ scattering, and $(M_{GT})_{\lambda}$ 
the Gamow-Teller matrix element for the beta-decay feeding of the resonance.

In the limit of a single narrow resonance, $\gamma_{\lambda}^2 P_2
\rightarrow 0$, then

\be
g(E_x) \rightarrow (M_{GT})_{\lambda}^2 \pi \delta(E_{\lambda} - E)
\label{narrowlimit}
\ee

\noindent and the rate expression, Eq.\ (\ref{rate}), reduces 
to the standard
one.  To obtain the electron spectrum, we integrate Eq.\ (\ref{rate})
over neutrino energies

\be
\frac{d \Gamma}{d E_e} \propto p_e E_e F( \pm Z,E_e) \frac{1}{\pi}
\int_{0}^{Q_{ec}+m-E_e} dE_x g(E_x) (Q_{ec}+m-E_x-E_e)^2 ,
\label{Ree}
\ee

\noindent while for the neutrino spectrum, we integrate over electron
energies

\be
\frac{d \Gamma}{d E_{\nu}} \propto E_{\nu}^2 \frac{1}{\pi}
\int_{0}^{Q_{ec}-E_{\nu}} dE_x g(E_x) \beta (Q_{ec}+m-E_x-E_{\nu})^2 
F( \pm Z, Q_{ec}+m-E_x-E_{\nu}) ,
\label{Rnu}
\ee

\noindent where $\beta = p_e/E_e = \left [ 1 - m^2/(Q_{ec}+m-E_x-E_{\nu})^2
\right ]^{1/2}$.

It remains to specify the parameters, which we take from the work of
Barker \cite{Ba89}, where fits are made simultaneously to the
$\alpha + \alpha$ scattering $d$-wave phase shift as well as the
beta-delayed $\alpha$-spectra.  Separate fits, however, are done
for the $^{8}$B and $^{8}$Li beta-delayed $\alpha$-spectra.  Thus there
are differences in the parameter sets between the two cases.  However,
these differences are small and can be attributable to isospin-symmetry 
breaking.  The adjustable parameters in the formulae include the
channel radius $a_2$, the boundary condition parameter $B_2$,
and the eigenenergies $E_{\lambda}$, reduced width amplitudes
$\gamma_{\lambda}$ and Gamow-Teller matrix elements $(M_{GT})_{\lambda}$
for the various $2^{+}$ levels, $\lambda$.  The best fits were
obtained with a large channel radius of $a_2 \simeq 6.5$ fm.   
For this value of $a_2$, the second $2^{+}$ level is at about 9 MeV, with a
width of about 10 MeV.  To date, there is no evidence for a state
in this vicinity.  Altogether five resonances are included, denoted
$\lambda = 1,2,3,0,1^{\prime}$.  The state, $\lambda = 1$, is the
dominant resonance corresponding to the first excited state in $^{8}$Be
at about 3 MeV; while $\lambda = 2$ is the second $2^{+}$ state
just mentioned, and $\lambda = 3$ represents a background state well
above the energy range being fitted and which naturally is not fed in 
beta decay, $(M_{GT})_3 = 0$.  In addition, there are two narrow states
at $E_x = 16.6$ and 16.9 MeV, 
which are isospin mixtures of $T=0$ and $T=1$ ~$2^{+}$
states, the latter being the analogue of the ground states of $^{8}$Li
and $^{8}$B and is fed by Fermi beta decay, $(M_{F})_{1^{\prime}} =
\sqrt{2}$.  The method of handling this isospin mixing is described
in Barker \cite{Ba69} and Warburton \cite{Wa86}.

The fitted parameters 
$E_{\lambda}$, $\gamma_{\lambda}$ and $(M_{GT})_{\lambda}$
depend on the choice $B_2$.  Identical fits can be found for different
choices; and connection formulae are available \cite{Ba72} to
relate one set of fitted parameters to another.  The standard choice is to
set $B_2 = S_2(E_{\mu})$, where $\mu$ is one of the resonances of the
fit.  The parameters obtained with this choice are 
labelled $E_{\lambda}^{(\mu)}$,
$\gamma_{\lambda}^{(\mu)}$ and $(M_{GT})_{\lambda}^{(\mu)}$.  
Barker \cite{Ba89}
only gives the parameter values for the case $\mu = \lambda$, and the
connection formulae have to be used to relate them all to a common $B_2$ 
value.  This is cumbersome to use; for our purposes it is sufficiently 
accurate to take the parameter values for $\mu = \lambda$ and to vary
$B_2$ as follows:  for $E_x < 5$ MeV where the $\lambda =1$ resonance 
dominates, use $B_2 = S_2(E_1)$; for
$5 < E_x < 13$ MeV where the $\lambda = 2$ resonance dominates, 
use $B_2 = S_2(E_2)$; and for $E_x > 13$ MeV 
where the doublet states dominate, use
$B_2 = S_2(\overline{E})$, where $\overline{E}$ is the average energy
of the two isospin-mixed states.  The parameter values are given in 
Table \ref{t:2}. 

\section{Recoil corrections}
\label{s:ARc}

Recoil corrections to allowed beta decay have been given by 
Holstein\cite{Ho74} and used by Bahcall and Holstein\cite{BH86}
in their discussion of corrections to the spectrum of neutrinos produced
by the beta decay of $^{8}$B in the sun.  There are two sources
to the recoil corrections: (a) true kinematic corrections
arising from relaxing the approximation that the recoiling nucleus
is at rest, and (b) the introduction of induced terms into the
V-A weak hadronic current, mainly the induced tensor term in the
vector current (weak magnetism) and the induced pseudotensor 
term in the axial current.  We have not
included an induced scalar term in the vector current, as it
gives a zero contribution under the assumption of the conserved vector
current (CVC) hypothesis, or an induced pseudoscalar term in the axial
current as it is small in beta decay.  The beta decay spectrum, 
therefore, is predominantly given by four form factors, denoted in the
formalism of Holstein \cite{Ho74} by $a$, $b$, $c$ and $d$ for the
Fermi, weak magnetism, Gamow-Teller and induced pseudotensor current form
factors respectively.

Let $P$, $p_r$, $p_e$ and $p_{\nu}$ denote the respective four momenta of
the parent nucleus, daughter nucleus, electron and neutrino.  Further,
let $M$ be the mass of the parent nucleus, $M_N(^8$Li) or $M_N(^8$B);
$M^{\prime}$ be the ground-state mass of the daughter nucleus, $M_N(^8$Be);
$E_x$ the excitation energy above the ground state; and $m$ the
mass of the electron\footnote{Holstein \cite{Ho74} writes $M_1$ and $M_2$
as the masses of the parent and daughter nuclei 
respectively, and $M$ as their
arithmetic mean.}. Then we define

\bea
q & = & P - p_r = p_e + p_{\nu},
\nonumber \\
\Delta & = & M - M^{\prime} - E_x.
\label{kin1}
\eea

\noindent Further the maximum electron energy is

\bea
A_0 & = & (M^2 +m^2 -M^{{\prime}2}) / 2 M   
\nonumber \\
& \simeq & \Delta - \frac{\Delta^2 - m^2}{2 M} ,
\label{Eemax}
\eea

\noindent and the maximum neutrino energy is

\bea
C_0 & = & \left [ M^2 - (m + M^{\prime})^2 \right ] / 2 M ,
\nonumber \\
C_0 + m & \simeq & \Delta - \frac{(\Delta - m)^2}{2 M} .
\label{Enumax}
\eea

\noindent The form factors are functions of the 
four-momentum transfer squared.
It is convenient to expand these form factors

\bea
a(q^2) & = & a_1 + a_2 (q^2/M^2) + \ldots ,
\nonumber \\
c(q^2) & = & c_1 + c_2 (q^2/M^2) + \ldots ,
\nonumber \\
b(q^2) & = & b + \ldots ,
\nonumber \\
d(q^2) & = & d + \ldots
\label{ffq2}
\eea

\noindent and retain just the terms shown.  In the formulae that follow, we
will drop the dependence on the four-momentum transfer, \ie \ neglect $a_2$
and $c_2$, and not display the electromagnetic corrections, which are of
order $(\alpha Z)$.  Both these effects are very small on the recoil 
corrections, although they are retained in the final computations.

The probability that an electron of energy, $E_e$, is produced in an
allowed transition in beta decay is proportional to

\be
\frac{d \Gamma}{d E_e} \propto (a_1^2 + c_1^2) F(\pm Z,E_e)
(\Delta - E_e)^2 p_e E_e r_e(E_e) ,
\label{Ratee1}
\ee

\noindent where $r_e(E_e)$ is the recoil correction, $F(\pm Z,E_e)$ is the
usual Fermi function and $p_e$ is the electron momentum,
$p_e = [E_e^2 - m^2]^{1/2}$.  The upper sign is for electron emission in
$\beta^{-}$ decay, the lower sign for positron emission in $\beta^{+}$
decay.  The recoil correction is

\bea
r_e(E_e) & = & \Biggl \{ 1 - \frac{2}{3} \frac{\Delta}{M} \frac{(c_1^2 + c_1d
\pm c_1b)}{(a_1^2 + c_1^2)} + \frac{2}{3} \frac{E_e}{M} \frac{(3a_1^2
+5c_1^2 \pm 2 c_1b)}{(a_1^2 + c_1^2)}
\nonumber \\[3mm]
& & - \frac{1}{3} \frac{m^2}{M E_e} \frac{(2c_1^2 + c_1d \pm 2c_1b)}
{(a_1^2 + c_1^2)} \Biggr \} \frac{(A_0-E_e)^2}{(\Delta -E_e)^2} .
\label{recoile}
\eea

\noindent This result has been given before by Holstein \cite{Ho74}.

For the neutrino spectrum, the probability that a neutrino of energy,
$E_{\nu}$, is produced in an allowed transition in beta decay is
proportional to

\be
\frac{d \Gamma}{d E_{\nu}} \propto (a_1^2 + c_1^2) F(\pm Z,\Delta - E_{\nu})
E_{\nu}^2 (\Delta - E_{\nu}) [(\Delta - E_{\nu})^2 - m^2 ]^{1/2}
r_{\nu}(E_{\nu}),
\label{Ratenu1}
\ee

\noindent where $r_{\nu}(E_{\nu})$ is the recoil correction given by

\bea
r_{\nu}(E_{\nu}) & = & \Biggl \{ 1 - \frac{m \Delta}{M} \left [ \frac{1}
{\Delta - E_{\nu}} + \frac{1}{\Delta -E_{\nu} + m} \right ]
\nonumber \\[3mm]
& & - \frac{2}{3} \frac{\Delta}{M} \frac{(c_1^2 + c_1d \mp c_1b)}
{(a_1^2 + c_1^2)}
+ \frac{2}{3} \frac{E_{\nu}}{M} \frac{(3a_1^2+5c_1^2 \mp 2c_1b)}
{(a_1^2 + c_1^2)}
\nonumber \\[3mm]
& & + \frac{1}{3} \frac{m^2}{M (\Delta - E_{\nu})} \frac{(2c_1^2-c_1d \mp
2 c_1 b)}
{(a_1^2 + c_1^2)} \Biggr \}
\frac{(C_0+ m - E_{\nu})[(C_0+m-E_{\nu})^2 - m^2]^{1/2}}
{(\Delta - E_{\nu})[(\Delta -E_{\nu})^2 - m^2]^{1/2}} .
\label{recoilnu}
\eea

\noindent Note that $r_{\nu}(E_{\nu})$ is not given by
$r_e(E_e \rightarrow \Delta - E_{\nu})$. The reason is that, in recoil
order, the energy available is distributed three ways:  to the electron,
to the neutrino and to the recoiling nucleus.  The replacement
$E_e \rightarrow \Delta - E_{\nu}$ is only correct in the approximation
that the recoiling nucleus is at rest.

It remains to specify the values for the form factors.  For the dominant 
transition in the decay of $^{8}$B or $^{8}$Li, $2^{+},T=1 \rightarrow
2^{+},T=0$, the transition is pure Gamow-Teller and $a_1 = 0$.  Then,
for the recoil correction it is only necessary to supply the ratios
$b/c_1$ and $d/c_1$.  For these we use the same values as Bahcall and
Holstein \cite{BH86}:

\bea
\frac{b}{A c_1} & = & 7.7 \pm 1.0 ,
\nonumber \\
\frac{d}{A c_1} & = & 1.9 \pm 1.3 ,
\label{bdvalu}
\eea

\noindent where $A$ is the mass number, $A=8$.  The weak magnetism value is
determined from the CVC hypothesis and the 
measurement of the M1 width of the
$2^{+}$ analogue state in $^{8}$Be \cite{Na75}.  This value 
is consistent with, but less precise than, the recent 
measurement of the M1 width of
de Braeckeler \etal \cite{DeB94}. The induced pseudotensor 
form-factor value is determined from fits to the measured
$\beta {\mbox{-}} \alpha$ correlations on
$^{8}$B and $^{8}$Li \cite{TG75}.  The form factor $d$ is the 
lesser precisely 
known of the two listed in Eq.\ (\ref{bdvalu}).  To sample the
dependence on $d$, some
results will be given for $d=0$.

\section{Radiative corrections}
\label{s:ARdc}

Radiative corrections to the electron spectrum from allowed beta decay have
been considered in a number of papers \cite{KS59,Si67,Si78,YSM73,YM76}.
In obtaining the corrections, integrations are carried out over the
allowed neutrino and photon energies and the results exhibited as a 
differential spectrum in electron energy.  The contributions to the
radiative corrections have two components: the emission of real
photons (internal bremsstrahlung) and virtual radiative corrections due, for 
example, to the exchange of photons between charged particles.
The differential rate for real photon emission in an allowed beta
transition is given by the expression \cite{BS95}

\bea
d \Gamma & \propto & \frac{\alpha}{2 \pi} \frac{1}{\epsilon}
Q^2 dQ E_{\nu}^2 dE_{\nu} p_e E_e dE_e dx \delta (\Delta - E_e -E_{\nu}
- \epsilon )
\nonumber \\
& & \times \left [ \left ( \frac{E_e + \epsilon }{E_e} \right )
\frac{\beta^2(1 - x^2 Q^2/\epsilon^2 )}{(\epsilon - \beta Q x)^2}
+ \frac{\epsilon}{E_e^2 (\epsilon - \beta Q x)} \right ].
\label{tripledf}
\eea

\noindent Here $\epsilon$ represents the photon energy,
$\epsilon = [ Q^2 + \lambda^2 ]^{1/2}$, $Q$ being the photon momentum
and $\lambda$ a small nonzero photon mass introduced to regulate the
infrared divergence.  Further, $E_e$ is the electron energy, $p_e$
the electron momentum, $\beta = p_e/E_e$, $\Delta$ the maximum electron
energy and $x$ the cosine of the angle between the electron and photon
directions.  The delta function is used to integrate over the neutrino 
energies.  The integrations over $Q$ and $x$ are however very delicate: 
the logarithmic pole in $\lambda$ has to be extracted to cancel with the 
$\lambda$ dependence coming from the virtual corrections.

In the calibration experiment at SNO, a $^{8}$Li source will be placed
inside the SNO detector.  The real internal 
bremsstrahlung photons emitted by
this source will in principle be detected.  So we cannot follow the normal 
procedure of obtaining the radiative correction by integrating over $Q$.
Suppose that SNO detects photons of energy greater than some threshold,
say $\omega$.  Then for $Q$ less than $\omega$, 
the normal procedure can be followed 
except that in the integration over $Q$, the upper limit $Q_{\rm max}$
is taken to be the lesser of $\omega$ or $\Delta - E_e$, rather than just
$\Delta - E_e$.  The result for real photon emission is

\bea  
\frac{d \Gamma}{dE_e} & \propto & (\Delta - E_e)^2 p_e E_e
\frac{\alpha}{\pi} g_{<}(E_e,\Delta )
\nonumber \\
g_{<}(E_e,\Delta) & = & 
\frac{1}{E_e^2 y^2} \left [ \frac{1}{2} y^2 Q_{\rm max}^2 - \frac{2}{3}
y Q_{\rm max}^3 + \frac{1}{4} Q_{\rm max}^4 \right ]
\frac{1}{2 \beta} \ln \left ( \frac{1+\beta }{1-\beta} \right )
\nonumber \\
& & + \frac{2}{y^2 E_e} \left [ (y^2 - 2 y E_e) Q_{\rm max}
+ \frac{1}{2} ( E_e - 2y) Q_{\rm max}^2 + \frac{1}{3} Q_{\rm max}^3 \right ]
\left [ \frac{1}{2 \beta} \ln \left ( \frac{1 + \beta}{1 - \beta} \right )
- 1 \right ] 
\nonumber \\
& & + 2 \ln \left ( \frac{Q_{\rm max}}{\lambda} \right )
\left [ \frac{1}{2 \beta} \ln \left (\frac{1 + \beta}{1 - \beta} \right ) 
-1 \right ]
+ {\cal C} ,
\label{realle}
\eea

\noindent where $y$ is $(\Delta - E_e)$, and $Q_{\rm max}$ the lesser of
$\omega$ or $y$.  The value of ${\cal C}$ is given by Kinoshita and 
Sirlin \cite{KS59}

\bea
{\cal C} & = & 2 \ln 2 \left [ \frac{1}{2 \beta} \ln \left ( \frac{1+\beta}
{1-\beta} \right ) - 1 \right ] + 1 + \frac{1}{4 \beta} \ln
\left ( \frac{1+\beta}{1-\beta} \right ) \left [ 2 + \ln \left (
\frac{1 - \beta^2}{4} \right ) \right ]
\nonumber \\
& & + \frac{1}{\beta} \left [ L(\beta ) - L(- \beta ) \right ]
+ \frac{1}{2 \beta} \left [ L \left (\frac{1 - \beta}{2} \right )
- L \left ( \frac{1 + \beta }{2} \right ) \right ] ,
\label{bigC}
\eea

\noindent where $L(z)$ is a Spence function

\be
L(z) = \int_{0}^{z} \frac{dt}{t} \ln ( \mid 1 - t \mid ) .
\label{Spence}
\ee

For the case when the photon energy exceeds $\omega$ we must proceed a
little differently.  We assume, for simplicity, that photons and electrons
are detected with equal efficiency and that it is the sum of the
energy deposited that is observed in the SNO detector.  Thus in
Eq.\ (\ref{tripledf}) we change variables from $dE_e dQ$ to
$dE_e dX$, where $X$ is the sum $E_e + Q$, and integrate over $E_e$
from $m$ to $X$.  The result is

\bea
\frac{d \Gamma }{d X} & \propto & (\Delta - X)^2 \beta (X) X^2 
\frac{\alpha}{\pi} g_{>}(X, \Delta ) ,
\nonumber \\
g_{>}(X, \Delta) & = & \frac{2}{\beta (X) X} F(X) \ln \left (
\frac{X-m}{\lambda} \right ) - \frac{2}{\beta (X) X} \int_{m}^{X}
dE \frac{F(X) - F(E)}{X-E}
\nonumber \\
& & + \frac{1}{2} \frac{(X J_1 - J_2)}{\beta (X) X^2} ,
\label{realgt}
\eea

\noindent where

\bea
J_1 & = & \int_{m}^{X} dE \ln \left ( \frac{1+\beta }{1-\beta } \right ),
\nonumber \\
J_2 & = & \int_{m}^{X} dE E \ln \left ( \frac{1+\beta }{1-\beta } \right ),
\nonumber \\
F(E) & = & \beta E \left [ \frac{1}{2 \beta} \ln \left ( \frac{1+\beta }
{1- \beta } \right ) - 1 \right ]
\label{J1J2}
\eea

\noindent and $\beta \equiv \beta (E) = (1 - m^2/E^2 )^{1/2}$.  To these
expressions must be added the contribution from virtual radiative
corrections, which have been given by Yokoo \etal \cite{YSM73}.
The result is

\bea
\frac{d \Gamma }{d E_e} & \propto & (\Delta - E_e)^2 p_e E_e
\frac{\alpha}{\pi} g_{v}(E_e,\Delta) ,
\nonumber \\
g_{v}(E_e, \Delta ) & = & {\cal A} + 3 \ln \left ( \frac{\Lambda}{M} \right )
+ \frac{3}{4},
\nonumber \\
{\cal A} & = & \frac{1}{2} \beta \ln \left ( \frac{1+\beta }{1-\beta } \right )
- 1 + 2 \ln \left ( \frac{\lambda }{m} \right ) \left [ \frac{1}{2 \beta}
\ln \left ( \frac{1+\beta }{1- \beta } \right ) - 1 \right ],
\nonumber \\
& & + \frac{3}{2} \ln \left ( \frac{M}{m} \right ) - \frac{1}{\beta }
\left [ \frac{1}{2} \ln \left ( \frac{1+\beta }{1-\beta } \right )
\right ]^2 + \frac{1}{\beta } L \left ( \frac{2 \beta }{1 + \beta } \right )
\label{virtual}
\eea

\noindent where $M$ is the nucleon mass, and $\Lambda$ a renormalization
scale.  In the early years, Yokoo \etal \cite{YSM73} invoked an
intermediate vector-boson model to argue that $\Lambda$ should be of the
order of the nucleon mass.  Following the development of the
Weinberg-Salam Standard Model, the virtual radiative correction includes
additionally the exchange of $Z$-bosons.  Sirlin \cite{Si78} has
shown, remarkably, that the form of the expression, Eq.\ (\ref{virtual}),
remains unchanged except that $\Lambda$ is now replaced by the mass of the
$Z$-boson, $m_Z$.  For our purposes, the value of $\Lambda$ is not
important as it only enters in a constant, energy-independent term and would
be absorbed into the normalization of the spectra.  In Fig.\
\ref{f:2} we used $\Lambda$ equal to the nucleon mass, $M$.  
Note that the infrared
divergence term in $\ln (\lambda )$ exactly cancels between the real
and virtual photon expressions, as it should.  In summary, the
radiative correction to the electron spectrum is

\bea
\frac{ d \Gamma }{d E} & \propto & (\Delta - E)^2 \beta (E) E^2
R_e(E, \Delta ),
\nonumber \\
R_e(E,\Delta ) & = & 1 + \frac{\alpha}{\pi} \left ( g_{<}(E,\Delta )
+ g_{v}(E,\Delta) \right ) ~~~~~~ {\rm for} ~~ E = E_e < \omega ,
\nonumber \\
& = & 1 + \frac{\alpha}{\pi} \left ( g_{>}(E,\Delta )
+ g_{v}(E,\Delta) \right ) ~~~~~~ {\rm for} ~~ E = X \geq \omega .
\label{Radce}
\eea

Finally, we consider the radiative correction for the neutrino spectrum.
This is not obtained from the electron spectrum radiative correction
by the substitution of $E_e \rightarrow \Delta - E_{\nu}$ because
the energy available to the leptons is distributed three ways:
to the electron, to the neutrino and to the internal bremsstrahlung photon.
This was first pointed out by Batkin and Sundaresan \cite{BS95}.
Since the neutrinos are originating from beta decays in the sun,
there is no question this time of the internal bremsstrahlung photons being
detected.  Thus, starting from Eq.\ (\ref{tripledf}) for the real photon
radiative correction, the delta function is used in the integration over
electron energies, while the integrations over $Q$ and $x$ are done
carefully to isolate the infrared singularity.  The result has been
given by Batkin and Sundaresan \cite{BS95}.  The virtual radiative 
correction is unchanged and given by Eq.\ (\ref{virtual}).  Putting
these together we get

\bea
\frac{d \Gamma }{d E_{\nu}} & \propto & E_{\nu}^2 (\Delta - E_{\nu} )^2
\beta R_{\nu}(E_{\nu},\Delta ) ,
\nonumber \\
R_{\nu}(E_{\nu},\Delta ) & = & 1 + \frac{\alpha }{\pi }
\left ( g_{\nu}(E_{\nu},\Delta ) + g_{v}(\Delta - E_{\nu}, \Delta )
\right ) ,
\nonumber \\
g_{\nu}(E_{\nu},\Delta ) & = & 2 \ln \left ( \frac{\Delta - E_{\nu} - m}
{\lambda } \right ) 
\left [ \frac{1}{2 \beta} \ln \left ( \frac{1+\beta }{1- \beta }
\right ) -1 \right ] + {\cal C}
+ \frac{1}{2 (\Delta - E_{\nu} )^2 \beta } I_1 + I_2 ,
\label{Radcnu}
\eea

\noindent where

\bea
I_1 & = & \int_{0}^{\Delta - E_{\nu} -m} dQ Q \ln \left ( \frac{1 +
\beta (Q)}{1 - \beta (Q)} \right ) ,
\nonumber \\
I_2 & = & \int_{0}^{\Delta - E_{\nu} -m} dQ \frac {Q}{2 \beta (\Delta -
E_{\nu})} \biggl [ \left [ (\Delta - E_{\nu} - Q)^2 - m^2 \right ]^{1/2}
F \left ( \beta (Q) \right ) 
\nonumber \\
& & ~~~~~~~~~~~~~~~~~~~~~~~~~~~~~~~~~~~~~ -
\left [ (\Delta - E_{\nu} )^2 - m^2 \right ]^{1/2}
F \left ( \beta (0) \right ) \biggr ] ,
\nonumber \\
F \left ( \beta (Q) \right ) & = & \frac{4}{Q^2} \left [ \frac{1}{2 \beta (Q)}
\ln \left ( \frac{1+\beta (Q)}{1-\beta (Q)} \right ) - 1 \right ] ,
\nonumber \\
\beta (Q) & = & \left [ 1 - \frac{m^2}{(\Delta - E_{\nu} - Q)^2}
\right ]^{1/2}
\label{I1I2}
\eea

\noindent
 and $\beta \equiv \beta (0)$.

\begin{table}
\begin{center}
\caption{Percentage $3 \sigma$ error in $\langle T_e \rangle_{\nu}$
from Bahcall and Lisi \protect\cite{BL96} and in the ratio,
$\langle T_e \rangle_{\nu} / \langle T_e \rangle_{e}$.
\label{t:1}}
\begin{tabular}{lcc}
 & & \\[-3mm]
\% Error in: & $\langle T_e \rangle_{\nu}$ &
$\langle T_e \rangle_{\nu} / \langle T_e \rangle_{e}$  \\
\tableline
 & & \\[-3mm]
Statistics of 5000 CC events & 0.98 & 0.98 \\
CC cross-section & 0.43 & 0.43 \\
Neutrino spectrum & 1.14 & 0.15 \\
Energy resolution & 0.94 & 0.11 \\
Absolute energy calibration & 2.04 & 0.41 \\[3mm]
Total error & 2.82 & 1.16\tablenotemark[1] \\
\end{tabular}
\vspace{-3mm}
\tablenotetext[1]{We stress this is an optimistic estimate, see text.
When explicit calibration measurements are completed this number could change.}
\end{center}
\end{table}

\begin{table}
\begin{center}
\caption{Parameters in the R-matrix fit of Barker \protect\cite{Ba89}
to the beta-delayed $\alpha$-spectra.
\label{t:2}}
\begin{tabular}{cdddcddd}
& & & & & & & \\[-3mm]
& \multicolumn{3}{c}{$^{8}$B decay} &
& \multicolumn{3}{c}{$^{8}$Li decay} \\
\cline{2-4}
\cline{6-8}
& & & & & & & \\[-3mm]
$\lambda$
&~$E_{\lambda}$\tablenotemark[1] & $\gamma_{\lambda}$\tablenotemark[2] &
$(M_{GT})_{\lambda}$ &
& $E_{\lambda}$\tablenotemark[1] & $\gamma_{\lambda}$\tablenotemark[2] &
$(M_{GT})_{\lambda}$ \\
\tableline
& & & & & & & \\[-3mm]
1 & 2.804 & 0.588 & 0.102 & & 2.798 & 0.591 & 0.108 \\
2 & 8.87 & 0.884 & $-0$.180 & & 8.85 & 0.880 & $-$0.181 \\
3 & 34.1 & 1.442 & 0.000 & & 34.1 & 1.442 & 0.000 \\
0 & 16.72 & 0.109 & 1.64 & & 16.72 & 0.109 & 1.77 \\
$1^{\prime}$ & 17.02 & \tablenotemark[3] & 0.000\tablenotemark[4] &
& 17.02 & \tablenotemark[3] & 0.000\tablenotemark[4] \\
\end{tabular}
\vspace{-3mm}
\tablenotetext[1]{This is the channel energy in MeV.}
\tablenotetext[2]{In units of MeV$^{1/2}$.}
\tablenotetext[3]{$\alpha$-decay from $T=1$ states is isospin forbidden.
However, two-state isospin mixing is included, see text.}
\tablenotetext[4]{However, a Fermi matrix element of $(M_F)_{1^{\prime}}
= \sqrt{2}$ is included.}
\end{center}
\end{table}

\newpage

\begin{figure}
\caption{Recoil corrections, Eqs.\ (\protect\ref{recoile}) and
(\protect\ref{recoilnu}), for the electron spectrum of $^{8}$Li
and the neutrino spectrum of $^{8}$B respectively, and
their ratio.  Graph (a) includes the induced 
pseudotensor current form factor;
graph(b) excludes the induced pseudotensor current form factor.
\label{f:1}}
\end{figure}

\begin{figure}
\caption{Radiative corrections, Eqs.\ (\protect\ref{Radce}) and
(\protect\ref{Radcnu}), for the electron spectrum of $^{8}$Li
and the neutrino spectrum of $^{8}$B beta decay respectively and
their ratio.  Graph (a) is the case when the internal bremsstrahlung  
photons are not detected, $\omega$ large (see Appendix \protect\ref{s:ARdc});
graph (b) is the case when
the internal bremsstrahlung photons are detected and the total energy
deposited, $X = E_e + Q$, is recorded in the SNO detector, $\omega$ small.
\label{f:2}}
\end{figure}

\begin{figure}
\caption{Values of the ratio,
$\langle T_e \rangle_{\nu} / \langle T_e \rangle_{e}$,
and their $3 \sigma$ errors due to various uncertainties.
Also shown are various theoretical values corresponding to different
neutrino-oscillation scenarios, taken from Bahcall and 
Lisi \protect\cite{BL96}.  
Labels: STD = standard (no oscillation); SMA = small-mixing angle (MSW);
LMA = large-mixing angle (MSW); VAC = vacuum oscillation.
\label{f:3}}
\end{figure}

\begin{figure}
\caption{
Ratios of normalized electron spectra as a function of the
electron kinetic energy, $T_e$ (see text).  The two
dashed lines enclose the error band without a $^8$Li calibration 
experiment, due to the $\pm 3 \sigma$ uncertainty in absolute
energy calibration.  The two solid lines enclose the error band 
in the ratio of measured $^8$B solar-neutrino to 
measured $^8$Li spectra under the same assumptions.
The data points are theoretical values
for the CC spectrum evaluated under the small-angle MSW solution (SMA).
\label{f:4}}
\end{figure}

\newpage
\setcounter{figure}{0}

\begin{figure}
\begin{minipage}[t]{7.5cm}
\epsfig{file=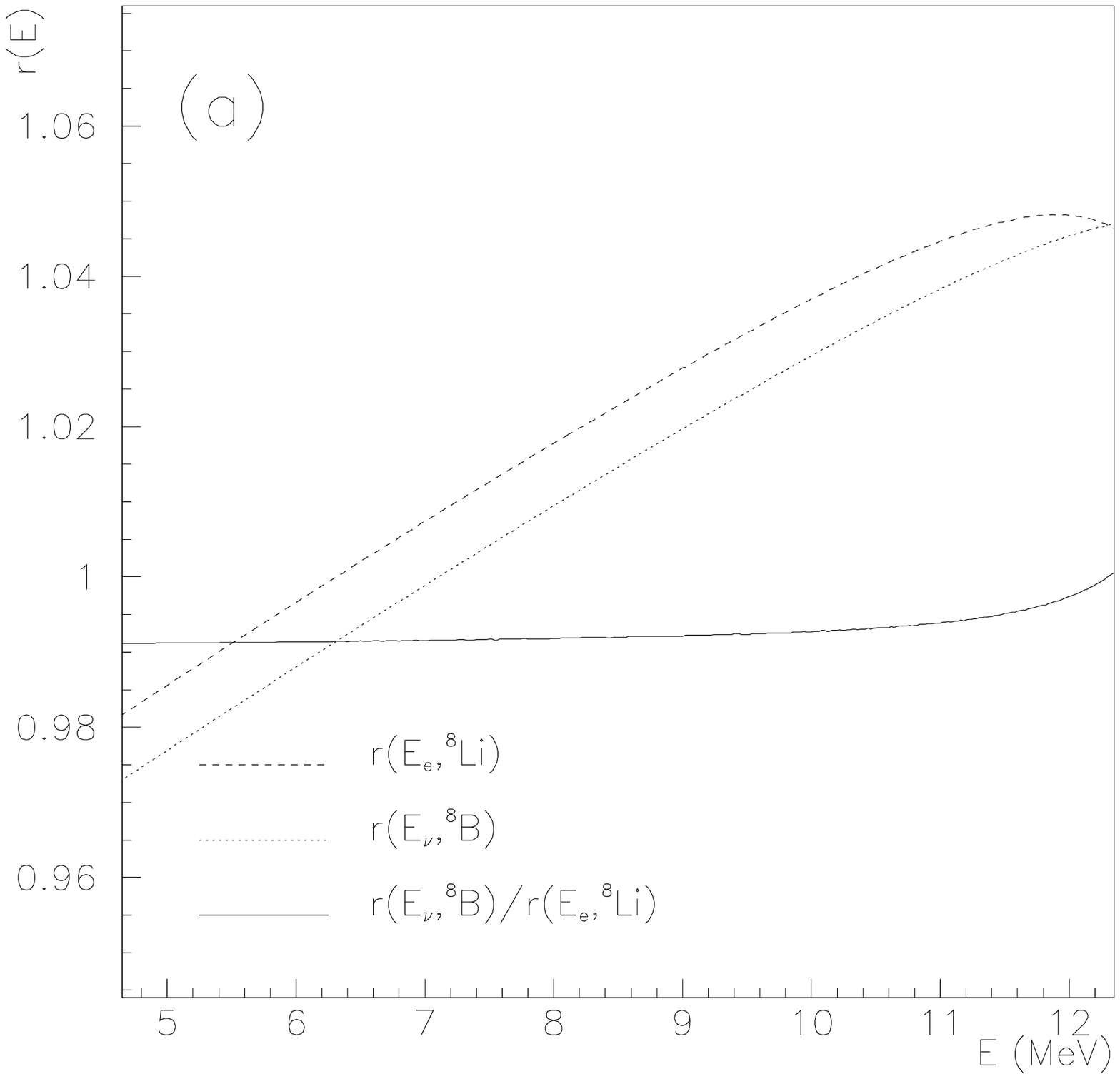,height=7cm}
\end{minipage}
\begin{minipage}[b]{7.5cm}
\epsfig{file=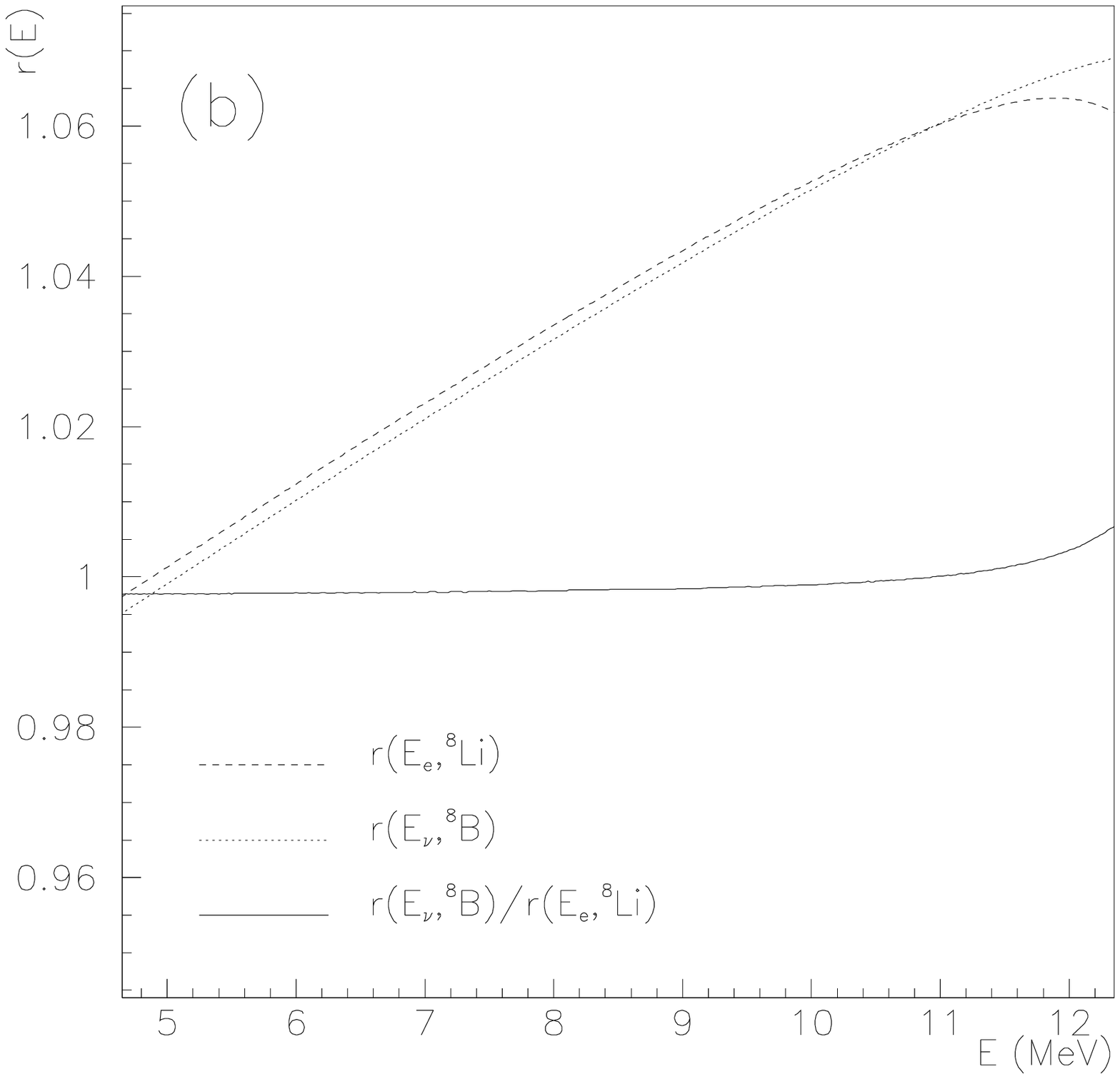,height=7cm}
\end{minipage}
\caption{ }
\end{figure}

\begin{figure}
\begin{minipage}[t]{7.5cm}
\epsfig{file=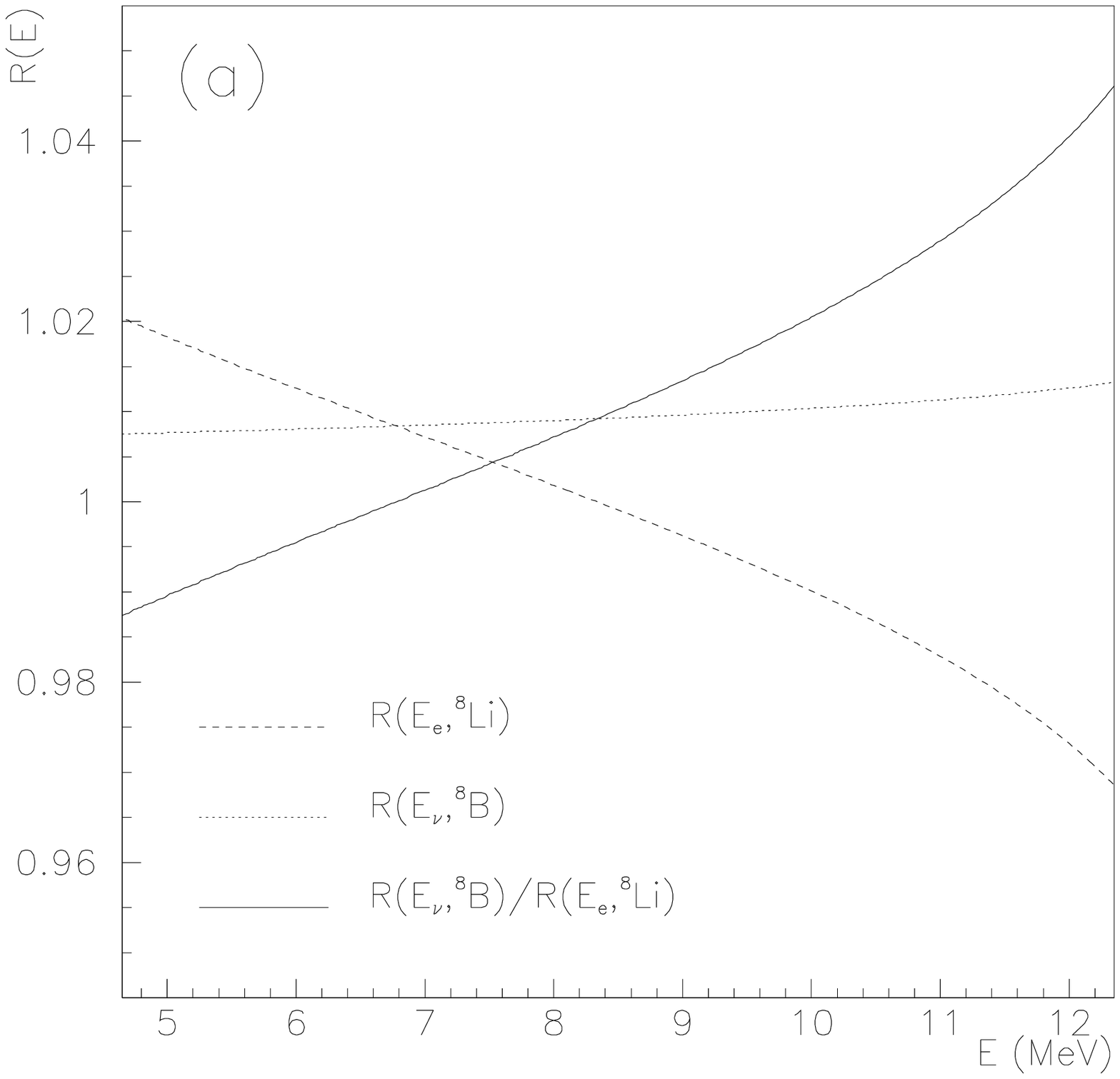,height=7cm}
\end{minipage}
\begin{minipage}[b]{7.5cm}
\epsfig{file=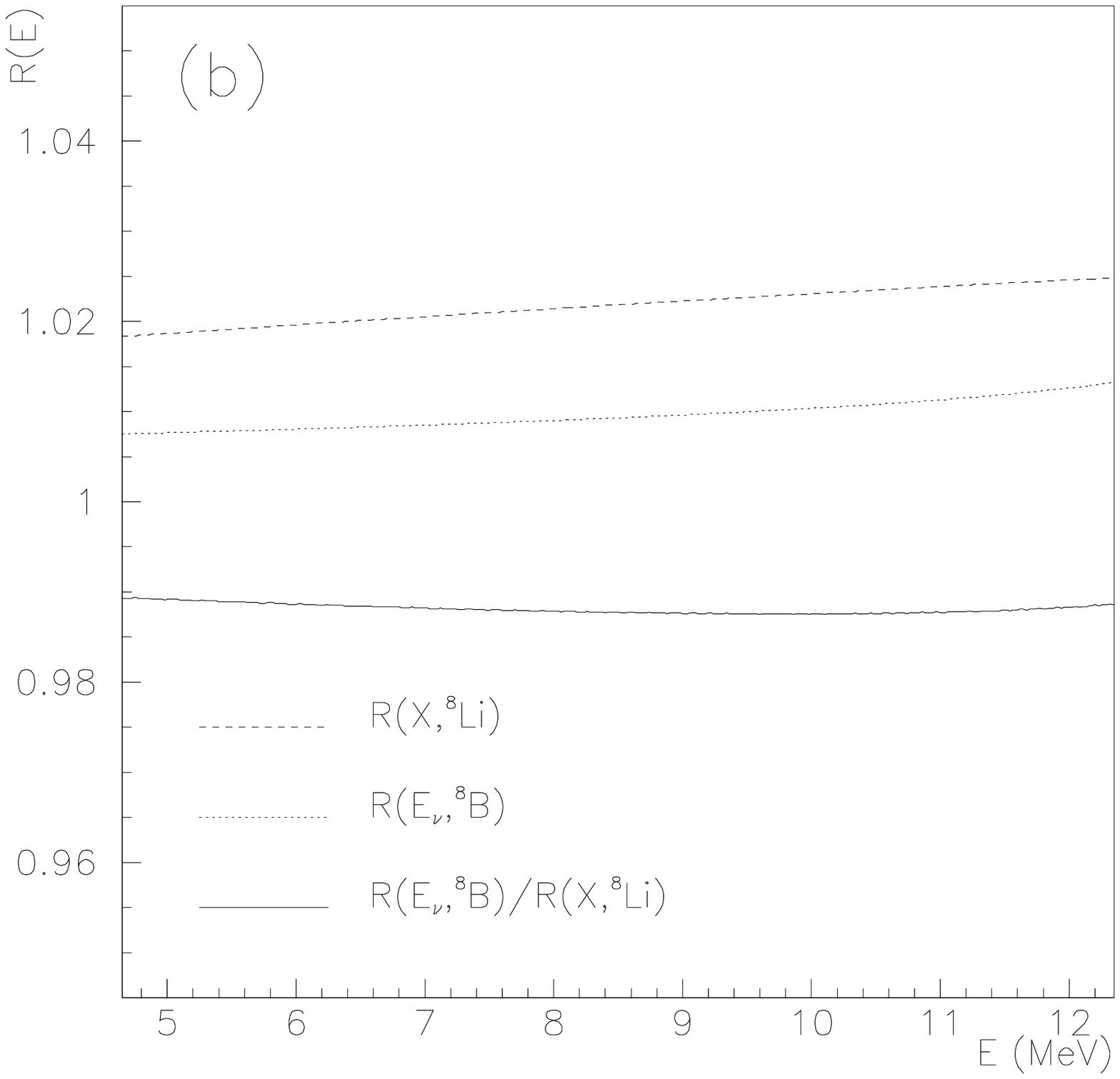,height=7cm}
\end{minipage}
\caption{ }
\end{figure}

\epsfysize=8cm
\begin{figure}
\begin{center}
\leavevmode
\epsffile{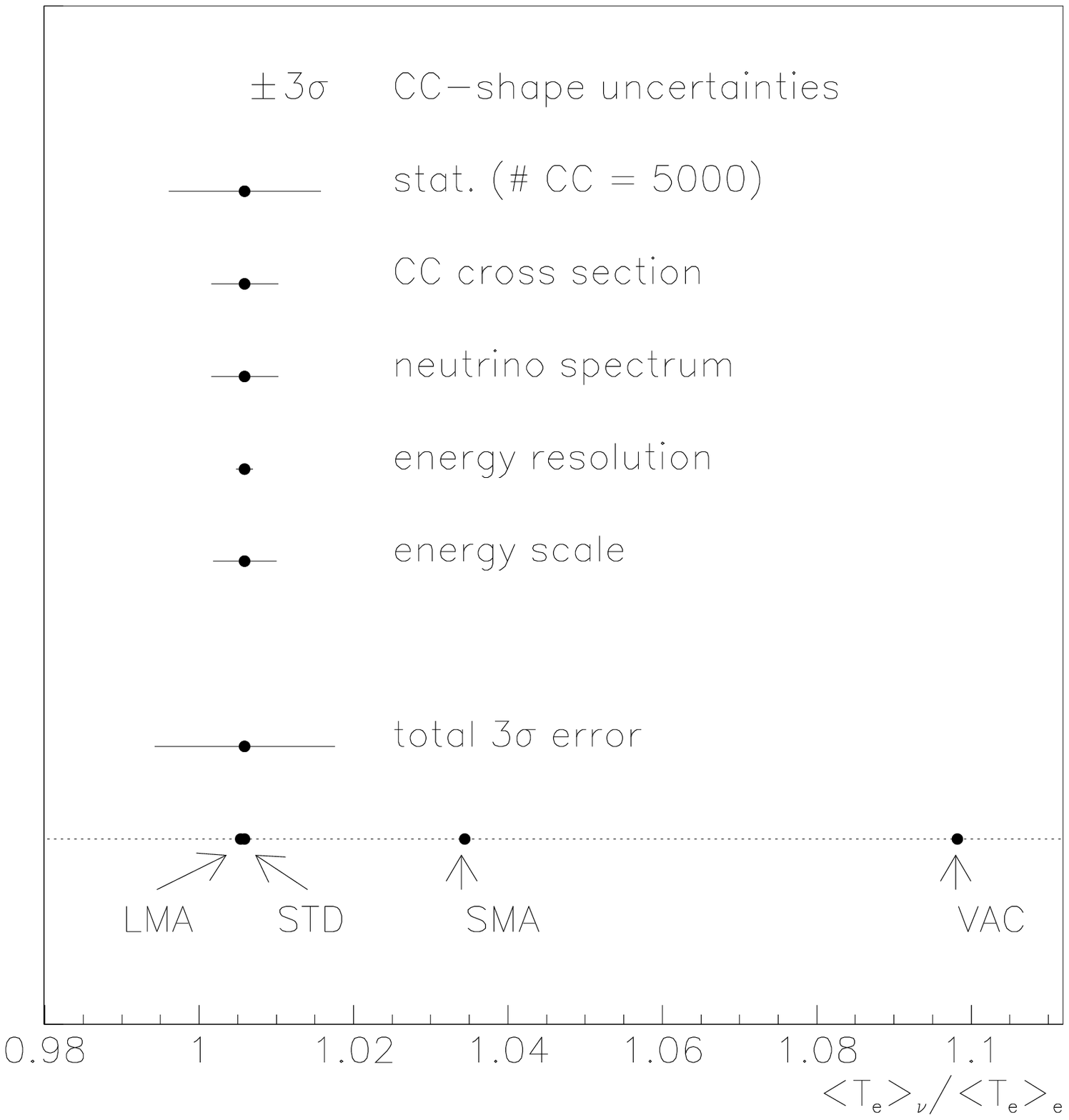}
\end{center}
\caption{ }
\end{figure}

\epsfysize=8cm
\begin{figure}
\begin{center}
\leavevmode
\epsffile{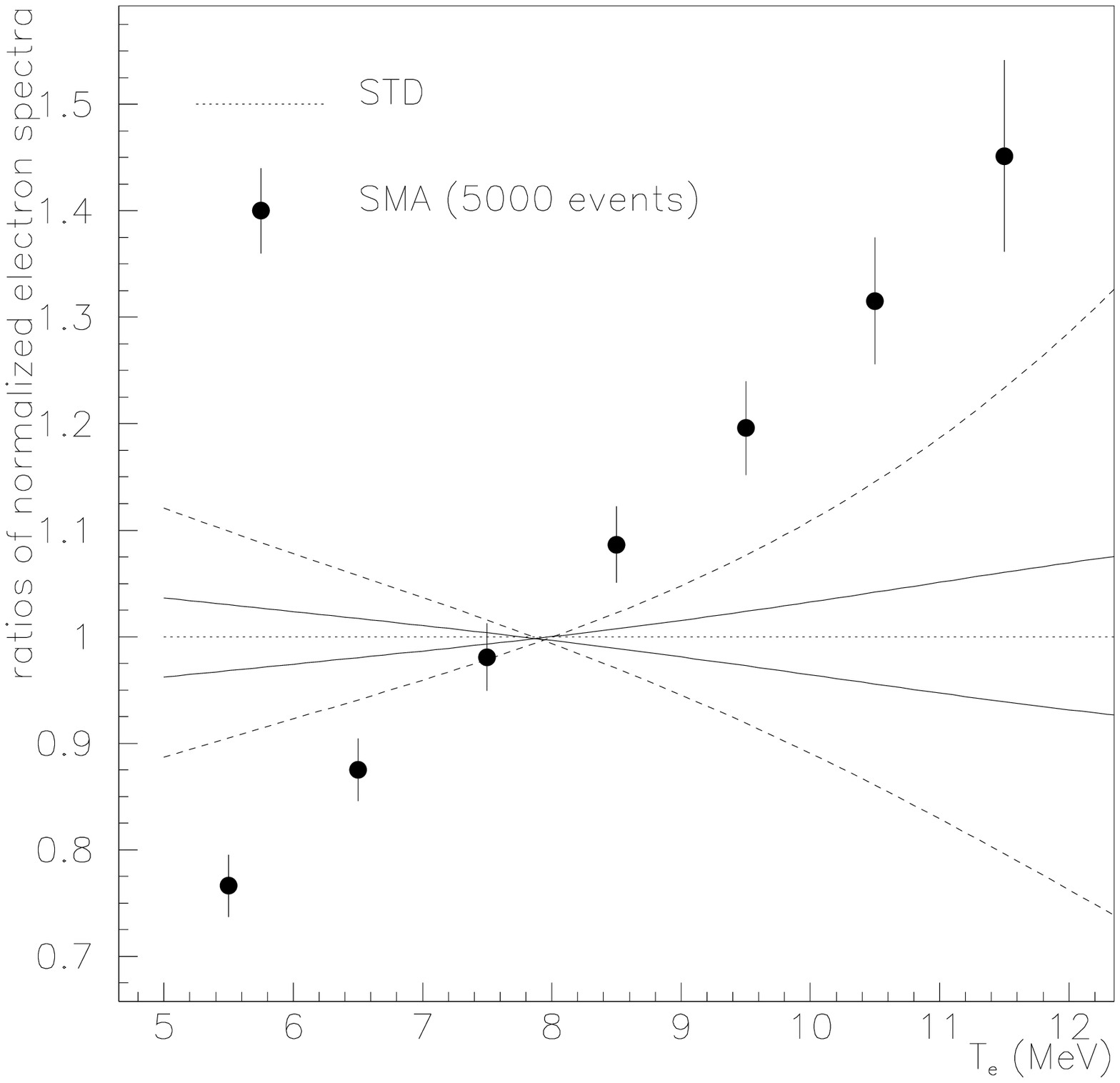}
\end{center}
\caption{ }
\end{figure}


\begin{references}
\bibitem{SNO}
SNO Collaboration, G.T. Ewan \etal , ``Sudbury Neutrino Observatory Proposal,"
Report No. SNO-87-12, 1987 (unpublished)

\bibitem{BA89}
J.N. Bahcall, {\it Neutrino Astrophysics} (Cambridge University Press,
Cambridge, England, 1989)

\bibitem{Su94}
B. Sur \etal , Bull. Am. Phys. Soc. {\bf 39}, 1389 (1994)

\bibitem{EB68}
S.D. Ellis and J.N. Bahcall, \np {\bf A114}, 636 (1968)

\bibitem{BL96}
J.N. Bahcall and E. Lisi, \pr {\bf D54}, 5417 (1996)

\bibitem{KN94}
K. Kubodera and S. Nozawa, Int. J. Mod. Phys. {\bf E3}, 101 (1994)

\bibitem{Expts}
D.H. Wilkinson and D.E. Alburger, \prl {\bf 26}, 1127 (1971);
B.J. Farmer and C.M. Class, \np {\bf 15}, 626 (1960);
L. De Braeckeleer and D. Wright, unpublished data, as quoted
in L. De Braeckeleer \etal , \pr {\bf C51}, 1767 (1995)

\bibitem{Ba96}
J.N. Bahcall, E. Lisi, D.E. Alburger, L. De Braeckeleer, S.J. Freedman,
and J. Napolitano, \pr {\bf C54}, 411 (1996)

\bibitem{Ba89}
F.C. Barker, Aust. J. Phys., {\bf 42}, 25 (1989)

\bibitem{Ho74} 
B.R. Holstein, Rev. Mod. Phys., {\bf 46}, 789 (1974)

\bibitem{BK96}
J.N. Bahcall and P.I. Krastev, \pr {\bf D53}, 4211 (1996)

\bibitem{Ba69}
F.C. Barker, Aust. J. Phys., {\bf 22}, 418 (1969)

\bibitem{Wa86}
E.K. Warburton, \pr {\bf C33}, 303 (1986)

\bibitem{Ba72}
F.C. Barker, Aust. J. Phys., {\bf 25}, 341 (1972)

\bibitem{BH86}
J.N. Bachall and B.R. Holstein, \pr {\bf C33}, 2121 (1986)

\bibitem{Na75}
A.M. Nathan \etal , \prl {\bf 35}, 1137 (1975); {\bf 49},
1056(E) (1982)

\bibitem{DeB94}
L. De Braeckeleer \etal , \pr {\bf C51}, 2778 (1995)

\bibitem{TG75}
R.E. Tribble and G.T. Garvey, \pr {\bf C12}, 967 (1975);
R.D. McKeown \etal , \pr {\bf C22}, 738 (1980)
 
\bibitem{KS59}
T. Konoshita and A. Sirlin, \pr {\bf 113}, 1652 (1959)

\bibitem{Si67}
A. Sirlin, \pr {\bf 164}, 1767 (1967)

\bibitem{Si78} 
A. Sirlin, Rev. Mod. Phys., {\bf 50}, 573 (1978)

\bibitem{YSM73} 
Y. Yokoo, S. Suzuki and M. Morita, Prog. Theor. Phys. {\bf 50}, 1894 (1973)

\bibitem{YM76}
Y. Yokoo and M. Morita, Suppl. Prog. Theor. Phys. {\bf 60}, 37 (1976)

\bibitem{BS95}
I.S. Batkin and M.K. Sundaresan, \pr {\bf D52}, 5362 (1995)

\end{references}
\end{document}